\documentclass[conference]{IEEEtran}
\usepackage{cite}

\ifCLASSINFOpdf
  \usepackage[pdftex]{graphicx}
\else
   \usepackage[dvips]{graphicx}
\fi

\usepackage[cmex10]{amsmath}
\usepackage{amssymb}
\usepackage{dsfont}

\usepackage{array}
\usepackage[ruled,boxed,linesnumbered]{algorithm2e}

\usepackage[caption=false, font=footnotesize]{subfig}

\usepackage{multirow}
\usepackage{threeparttable}
\usepackage{url}
\usepackage{stfloats} 
\begin{document}
%
\title{A Fast Decodable Full-Rate STBC with High Coding Gain  for $4\times2$ MIMO Systems}


\author{
\IEEEauthorblockN{Ming Liu, Maryline H\'elard, Jean-Fran\c{c}ois H\'elard, Matthieu Crussi\`ere}
\IEEEauthorblockA{Universit\'e Europ\'eenne de Bretagne (UEB)\\
INSA, IETR, UMR 6164, F-35708, Rennes, France \\
Email: \{ming.liu; maryline.helard;  jean-francois.helard; matthieu.crussiere\}@insa-rennes.fr}
}

\maketitle

\begin{abstract}
In this work, a new fast-decodable space-time block code (STBC) is proposed.
The code is full-rate and full-diversity for $4\times2$ multiple-input multiple-output (MIMO) transmission.
Due to the unique structure of the codeword, the proposed code requires a much lower computational complexity to provide maximum-likelihood (ML) decoding performance.
It is shown that the ML decoding complexity is only $O(M^{4.5})$ when $M$-ary square QAM constellation is used.
Finally, the proposed code has highest minimum determinant among the fast-decodable STBCs known in the literature. Simulation results prove that the proposed code provides the best bit error rate (BER) performance among the state-of-the-art STBCs.
\end{abstract}

\IEEEpeerreviewmaketitle

\section{Introduction}
Quasi-orthogonal space-time block codes (STBCs)~\cite{jafarkhani2001quasi} support full coding rate for the multiple-input multiple-output (MIMO) transmissions with more than two transmit antennas.
However, they suffer from high maximum-likelihood (ML) decoding complexity when the high order constellation is used.

The so-called fast decodable STBCs~\cite{biglieri2009fast,srinath2009low,srinath2011generalized,ismail2011new,ismail2013new} exploit the orthogonality embedded in the codewords and achieve ML decoding performance at a low decoding complexity.
For instance, Biglieri-Hong-Viterbo (BHV) constructed a rate-2 STBC~\cite{biglieri2009fast} based on the classical Jafarkhani code~\cite{jafarkhani2001quasi}. Exploiting the quasi-orthogonal structure  embedded in the Jafarkhani codeword, the BHV code requires much less complexity than the real ML decoding without losing performance.
Ismail-Fiorina-Sari (IFS) proposed a rate-1 STBC for four-transmit-antenna systems~\cite{ismail2011new} based on the rate-$3/4$ complex orthogonal design (COD)~\cite{tirkkonen2002square}.
Thanks to the underlying orthogonal structure, the ML decoding is achieved by using a low complexity conditional detector followed by hard decisions.
This construction idea was then generalized to rate-2 STBC case~\cite{ismail2013new}.
Srinath and Rajan (Srinath-Rajan) proposed a fast-decodable rate-2 STBC~\cite{srinath2009low,srinath2011generalized} based on co-ordinate interleaved orthogonal design (CIOD)~\cite{khan2006single}. The Srinath-Rajan code possesses high coding gain but needs low decoding complexity due to its orthogonal structure.
Similarly, some other STBCs such as DjABBA code~\cite{hottinen2004precoder}, Golden code~\cite{belfiore2005golden} and 3D MIMO code~\cite{nasser20083d} were shown to be
also fast-decodable because of the underlying orthogonality in the codewords.

In this paper, we investigate the full-rate STBC for four-transmit-two-receive-antenna ($4\times2$) MIMO transmission which is a typical asymmetric MIMO scenario considered in LTE, DVB-NGH~\cite{liu2013distributed} etc.
We propose a new fast-decodable STBC that requires low decoding complexity and possesses high coding gain.
The main contributions of the paper are:
\begin{itemize}
  \item A new STBC codeword that enables low-complexity MIMO decoding is proposed;
  \item Theoretical analysis proves that the new code is one of the least complex STBCs with full-rate for $4\times2$ MIMO systems;
  \item Corresponding low-complexity ML decoding method for the new code is presented;
  \item The optimal rotation angle is proposed for the new codeword, which leads to the highest coding gain;
  \item Simulation results prove that the new code provides the best BER performance.
\end{itemize}

The reminder of the paper is organized as follows. Section~\ref{sec:sys_mod} introduces the MIMO system model. The novel STBC is proposed in Section~\ref{sec:new_STBC}. Low-complexity decoding method is also presented  in this part.
Simulation results are given in Section~\ref{sec:simu}.
Conclusions are drawn in section~\ref{sec:conclusion}.

\emph{Notations}: $\mathbf{X}^T$ and $\mathbf{X}^{\mathcal{H}}$ denote the transpose and conjugate transpose of a matrix $\mathbf{X}$. $\mathbf{I}_n$ and $\mathbf{O}_n$ denote the identity matrix and null matrix of size $n$.
$x^R$ and $x^I$ represent the real and imaginary parts of a complex variable $x$.
The operator $(\check{\cdot})$ realizes the complex to real conversion:
\begin{equation}\label{eq:check_function}
    \check{x} \triangleq \left [\begin{array}{*{20}c}
      x^R & -x^I \\
      x^I & x^R \\
    \end{array}\right ].
\end{equation}
The operation $\widetilde{\mathbf{x}}\triangleq[x_1^R,x_1^I,\ldots, x_n^R,x_n^I]^T$ separates the real and imaginary parts of the complex vector $\mathbf{x}$.
The operation $vec(\mathbf{X})\triangleq[\mathbf{x}_1^T,\mathbf{x}_2^T,\ldots, \mathbf{x}_n^T]^T$ stacks the columns of $\mathbf{X}$, i.e. $\mathbf{x}_j$'s, one below another.
The combined operation
$\widetilde{vec(\mathbf{X})}\triangleq\widetilde{[{\mathbf{x}}_1^T,{\mathbf{x}}_2^T,\ldots, {\mathbf{x}}_n^T]^T}$ first converts matrix $\mathbf{X}$ into stacked vector and then separates the real and imaginary parts.
$\langle \mathbf{x},\mathbf{y}\rangle\triangleq\mathbf{x}^T\mathbf{y}$ denotes the inner product of two vectors $\mathbf{x}$ and $\mathbf{y}$. $\otimes$ represents the Kronecker product.

\section{System Model}
\label{sec:sys_mod}
\subsection{MIMO system model}
The codeword matrix of a linear dispersion STBC that contains $\kappa$ information symbols can be represented by a linear combination~\cite{srinath2009low}:
\begin{equation}
  \mathbf{X}=\sum_{j=1}^{\kappa}s_j^R\mathcal{A}_{2j-1} + s_j^I\mathcal{A}_{2j},
\end{equation}
where $\mathcal{A}_{2j-1} \in \mathbb{C}^{N_t\times T}$ ($\mathcal{A}_{2j} \in \mathbb{C}^{N_t\times T}$), $j=1,2,\ldots,\kappa$, are the complex weight matrices representing the contribution of the real (imaginary) part of the $j$th information symbol $s_j$ in the final codeword matrix.
One STBC codeword $\mathbf{X}\in\mathbb{C}^{N_t\times T}$ is transmitted by $N_t$ transmit antennas over $T$ channel uses.

Assuming that the receiver has $N_r$ receive antennas, the received signal can be expressed as:
\begin{equation}\label{eq:mimo_eq}
  \mathbf{Y}=\mathbf{H}\mathbf{X}+\mathbf{W},
\end{equation}
where $\mathbf{H}\in\mathbb{C}^{N_r\times N_t}$ is the MIMO channel matrix in which the ($j$,$k$)th element $h_{j,k}$ represents the gain of the channel link between the $k$th transmit antenna and $j$th receive antenna; $\mathbf{Y},\ \mathbf{W}\in\mathbb{C}^{N_r\times T}$ represent the received signal and noise, respectively.
The channel is assumed to be quasi-static, which is a common assumption that is guaranteed by the system design.
The MIMO transmission in (\ref{eq:mimo_eq}) can be rewritten in a real-valued equivalent form:
\begin{equation}\label{eq:rec_sig_real_eq}
    \widetilde{\mathbf{y}}=\mathbf{H}_{eq}\widetilde{\mathbf{s}}+\widetilde{\mathbf{w}},
\end{equation}
where $\widetilde{\mathbf{y}} = \widetilde{vec(\mathbf{Y})}$, $\widetilde{\mathbf{w}} = \widetilde{vec(\mathbf{W})}$ and $\mathbf{H}_{eq}\in \mathbb{R}^{2N_rT\times2\kappa}$ is the equivalent channel matrix and is computed by~\cite{biglieri2009fast}:
\begin{equation}\label{eq:Heq}
  \mathbf{H}_{eq}=(\mathbf{I}_{T}\otimes\check{\mathbf{H}})\mathbf{G},
\end{equation}
in which the generator matrix $\mathbf{G}\in \mathbb{R}^{2N_tT\times2\kappa}$ is obtained by:
\begin{equation}\label{eq:generator_matrix}
    \mathbf{G}\triangleq[\widetilde{vec(\mathcal{A}_1)}, \widetilde{vec(\mathcal{A}_2)},\ldots,\widetilde{vec(\mathcal{A}_{2\kappa})}].
\end{equation}

\subsection{ML detection using sphere decoder}
We represent the equivalent channel matrix in column vectors, i.e. $\mathbf{H}_{eq}\triangleq[\mathbf{h}_1,\mathbf{h}_2,\ldots,\mathbf{h}_{2\kappa}]$.
After QR decomposition, the equivalent channel matrix can be decomposed into an orthogonal matrix $\mathbf{Q}$ and an upper triangular matrix $\mathbf{R}$, i.e. $\mathbf{H}_{eq}=\mathbf{Q}\mathbf{R}$ where $\mathbf{Q}\triangleq[\mathbf{q}_1,\mathbf{q}_2,\ldots,\mathbf{q}_{2\kappa}]$ and
\begin{equation}
   \mathbf{R}\triangleq\left [\begin{array}{*{20}c}
  \|\mathbf{r}_1\|^2 & \langle \mathbf{q}_1,\mathbf{h}_2\rangle & \cdots & \langle \mathbf{q}_1,\mathbf{h}_{2\kappa}\rangle \\
  0 & \|\mathbf{r}_2\|^2 & \cdots & \langle \mathbf{q}_2,\mathbf{h}_{2\kappa}\rangle \\
  \vdots & \vdots & \ddots & \vdots \\
  0&0 &\cdots & \|\mathbf{r}_{2\kappa}\|^2 \\
 \end{array}\right],
\end{equation}
where $\mathbf{r}_1=\mathbf{h}_1$, $\mathbf{r}_{j}=\mathbf{h}_j-\sum_{k=1}^{j-1}\langle \mathbf{q}_k,\mathbf{h}_j\rangle \mathbf{q}_k$,
$\mathbf{q}_j=\mathbf{r}_j/\|\mathbf{r}_j\|$, $j=1,\ldots,2\kappa$.
From (\ref{eq:rec_sig_real_eq}) and taking advantage of QR decomposition of $\mathbf{H}_{eq}$, the ML solution of the transmitted symbols can be acquired by:
\begin{equation}\label{eq:SD_detection_metric}
    \hat{\mathbf{s}}=\arg\min_{\mathbf{s}\in\boldsymbol\Theta^\kappa}\|\widetilde{\mathbf{z}}-\mathbf{R}\widetilde{\mathbf{s}}\|^2,
\end{equation}
where $\mathbf{z}=\mathbf{Q}^T\mathbf{y}$ is a linear transform of received signal, and $\boldsymbol\Theta$ is the set of constellation symbols.
It is actually a joint search of $\kappa$ information symbols.
The resulting complexity is $O(M^\kappa)$ when the constellation is $M$-QAM.

\section{Proposed Fast-decodable STBC}
\label{sec:new_STBC}
\subsection{Codeword matrix of DjABBA code}

Let's recall the codeword matrix of the DjABBA code~\cite{hottinen2004precoder}:
\begin{equation}\label{eq:DjABBA_cw}
\textbf{X}= \left [ \begin{array}{*{20}c}
        \cos\rho\ \mathbf{X}_A + \sin\rho\ \mathbf{X}_C & \cos\rho\ \mathbf{X}_B + \sin\rho\ \mathbf{X}_D  \\
        i(\sin\rho\ \mathbf{X}_B - \cos\rho\ \mathbf{X}_D) & \sin\rho\ \mathbf{X}_A - \cos\rho\ \mathbf{X}_C\\
        \end{array} \right],
\end{equation}
where $\rho$ is the a rotation angle dedicated to optimizing the pairwise error of the code; $\mathbf{X}_A$, $\mathbf{X}_B$, $\mathbf{X}_C$ and $\mathbf{X}_D$ are four Alamouti codewords~\cite{alamouti1998simple} associated with four information symbol pairs $[s_1\ s_2]$, $[s_3\ s_4]$, $[s_5\ s_6]$ and $[s_7\ s_8]$, respectively. For instance,
\begin{equation}\label{eq:}
    \mathbf{X}_A = \left [ \begin{array}{*{20}c}
        s_1 & s_2  \\
        -s_2^{\ast} & s_1^{\ast}\\
        \end{array} \right].
\end{equation}
Since eight information symbols are transmitted over four channel uses, leading to a space-time coding rate of $2$ which means full-rate for two-receive-antenna systems.

\begin{figure}[!t]
\centering
\includegraphics[width=2.3in]{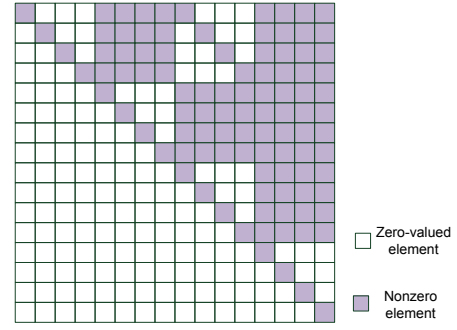}
\caption{$\mathbf{R}$ matrix of DjABBA code in quasi-static channel.}
\label{fig_R_mat}
\end{figure}
After performing QR decomposition to the channel matrix, the resulting upper triangular matrix $\mathbf{R}$ is illustrated in Fig.~\ref{fig_R_mat}.
The contributions of information symbols $s_1^R$, $s_1^I$, $s_2^R$ and $s_2^I$ are not correlated in the received signal
which could be exploited to achieve a ML decoding complexity of $O(M^6)$, instead of $O(M^8)$~\cite{srinath2009low}.
More interestingly, we also notice that this uncorrelation also exists in other information symbol pairs.
This motivates us to propose a new codeword that fully exploits the embedded orthogonality to achieve lower decoding complexity.

\subsection{Proposed code}

As mentioned previously, low decoding complexity is commonly achieved by exploiting the orthogonality between information symbols.
In the literature, many fast-decodable STBCs~\cite{biglieri2009fast,srinath2009low,ismail2011new} were designed so that some information symbols are uncorrelated with others. This symbol independency enables group-wise detections in parallel which can significantly reduce the decoding complexity.

As far as the DjABBA code is concerned, the orthogonality embedded in the Alamouti structure can be exploited to achieve lower decoding complexity.
For instance, as shown in Fig.~\ref{fig_R_mat},  the $4\times4$ submatrices located in the main diagonal positions of the $\mathbf{R}$ matrix are identity matrices. This means that the two information symbols in each Alamouti codeword (e.g. $\mathbf{X}_A$, $\mathbf{X}_B$, $\mathbf{X}_C$ and $\mathbf{X}_D$) are uncorrelated because of the orthogonality in the Alamouti structure.
The ML decoding complexity is reduced from $O(M^8)$ to $O(M^6)$ thanks to this orthogonality within each Alamouti codeword.
\emph{Intuitively, even lower complexity can be achieved if we can build orthogonality among more information symbols.}

Before proposing the new codeword, let's recall a simple fact related to the Alamouti structure.

\vspace{0.2cm}
\emph{Lemma 1}:  For two Alamouti codewords $\mathbf{A}$ and $\mathbf{B}$, its linear combination $\mathbf{C}=a\mathbf{A}+b\mathbf{B}$ satisfies $\mathbf{C}\mathbf{C}^{\mathcal{H}}=\mathbf{C}^{\mathcal{H}}\mathbf{C}=c\mathbf{I}_2$, where $a$, $b$ and $c$ are scalar numbers.
\vspace{0.1cm}

The lemma can be proved via some simple manipulations. Its proof is omitted here. This lemma actually means that the linear combination of two Alamouti codewords preserves the Alamouti structure as well as the orthogonality between different parts.
It gives a way to build orthogonality among more information symbols.

Moreover, it is known that the decoding complexity is mainly determined by the orthogonality among the first several information symbols when the conditional detection is used~\cite{biglieri2009fast,srinath2009low}. For example, in the design of BHV code, the low decoding complexity is  achieved by the Jafarkhani-like quasi-orthogonal structure embedded in the first four information symbols.
This motivates us to build orthogonality among $\{s_1,s_2,s_3,s_4\}$ in the DjABBA codeword in order to reduce the decoding complexity.

\begin{figure}[!t]
\centering
\includegraphics[width=2.3in]{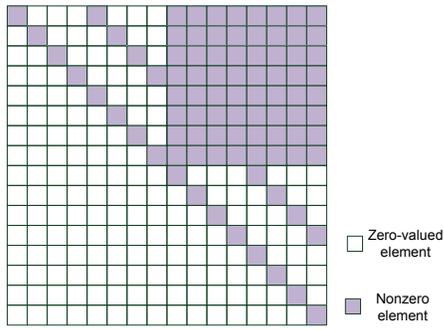}
\caption{$\mathbf{R}$ matrix of the new code in quasi-static channel.}
\label{fig_R_mat_new}
\end{figure}

With the knowledge presented above, we propose a new codeword as:
\begin{equation}\label{eq:newDjABBA}
\textbf{X}_{new}\! =\!\! \left [ \begin{array}{*{20}c}
        \cos\rho\ \mathbf{X}_A + \sin\rho\ \mathbf{X}_B & \cos\rho\ \mathbf{X}_C + \sin\rho\ \mathbf{X}_D  \\
        i(\sin\rho\ \mathbf{X}_C - \cos\rho\ \mathbf{X}_D) & \sin\rho\ \mathbf{X}_A - \cos\rho\ \mathbf{X}_B\\
        \end{array} \right],
\end{equation}
where the rotation angle $\rho$ is chosen to be $\rho = \tan^{-1}(\frac{1+\sqrt{5}}{2})$ in order to maximize the coding gain.
Some discussion on the selection of $\rho$ will be given later.

The new codeword is very similar to the original one presented in (\ref{eq:DjABBA_cw}) except that it is formed by the linear combinations of $\mathbf{X}_A$ and $\mathbf{X}_B$, as well as those of $\mathbf{X}_C$ and $\mathbf{X}_D$.
\newcounter{MYtempeqncnt1}
\begin{figure*}[!t]
\normalsize
\vspace*{4pt}
\setcounter{MYtempeqncnt1}{\value{equation}}
\setcounter{equation}{11}
\begin{align}
\label{eq:new_codeword}
\textbf{X}_{new}=\left [ \begin{array}{*{20}c}
        \cos\rho\ s_1 + \sin\rho\ s_3 & \cos\rho\ s_2 + \sin\rho\ s_4 & \cos\rho\ s_5 + \sin\rho\ s_7 & \cos\rho\ s_6 + \sin\rho\ s_8 \\
        -\cos\rho\ s_2^{\ast} - \sin\rho\ s_4^{\ast} & \cos\rho\ s_1^{\ast} + \sin\rho\ s_3^{\ast} & -\cos\rho\ s_6^{\ast} - \sin\rho\ s_8^{\ast} & \cos\rho\ s_5^{\ast} + \sin\rho\ s_7^{\ast} \\
        i(\sin\rho\ s_5 - \cos\rho\ s_7) & i(\sin\rho\ s_6 - \cos\rho\ s_8) & \sin\rho\ s_1 - \cos\rho\ s_3 & \sin\rho\ s_2 - \cos\rho\ s_4 \\
        -i(\sin\rho\ s_6^{\ast} - \cos\rho\ s_8^{\ast}) & i(\sin\rho\ s_5^{\ast} - \cos\rho\ s_7^{\ast}) & -\sin\rho\ s_2^{\ast} + \cos\rho\ s_4^{\ast} & \sin\rho\ s_1^{\ast} - \cos\rho\ s_3^{\ast} \\
        \end{array} \right]
\end{align}
\setcounter{equation}{12}
\hrulefill
\end{figure*}
These modifications, as will be shown later, yield orthogonality among more information symbols, and therefore lead to a decoding complexity reduction.
A detailed expression of the codeword matrix is given in (\ref{eq:new_codeword}) shown on next page.
After QR decomposition, the $\mathbf{R}$ matrix has a very good structure, as depicted in Fig.~\ref{fig_R_mat_new}.
More precisely, the property of $\mathbf{R}$ matrix can be expressed in the following theorem.

\vspace{0.2cm}
\emph{Theorem 1}: For $j,k\in\{1,2,\ldots,8\}$, in the upper triangular matrix $\mathbf{R}$  we have $ \langle \mathbf{q}_{j},\mathbf{h}_k\rangle=0,\ \forall k\neq j\ \mathrm{or}\ \forall k\neq j+4$.
\vspace{0.1cm}

\emph{Proof}:
From the codeword matrix (\ref{eq:new_codeword}), it can be easily verified that, for $j,k\in\{1,2,\ldots,8\}$:
\begin{equation}\label{eq:}
 \mathcal{A}_j\mathcal{A}_k^{\mathcal{H}}+\mathcal{A}_k\mathcal{A}_j^{\mathcal{H}}=\mathbf{O}_4,\quad \forall k\neq j\ \mathrm{or}\ \forall k\neq j+4.
\end{equation}
Using the Theorem 2 given in \cite{srinath2009low}, it yields that the corresponding $j$th and $k$th columns of equivalent channel matrix $\mathbf{H}_{eq}$ are orthogonal, i.e. for $j,k\in\{1,2,\ldots,8\}$:
\begin{equation}\label{eq:hjhk}
\langle \mathbf{h}_j,\mathbf{h}_k\rangle,\quad \forall k\neq j\ \mathrm{or}\ \forall k\neq j+4.
\end{equation}
Consequently, according to the definition of QR decomposition, it can be obtained that, for $j\in\{1,2,3,4\}$:
\begin{equation}\label{eq:rjqj}
\mathbf{r}_j=\mathbf{h}_j,\ \mathbf{q}_j=\mathbf{h}_j/\|\mathbf{h}_j\|.
\end{equation}
Using (\ref{eq:hjhk}) and (\ref{eq:rjqj}), it is obviously that, for $j\in\{1,2,3,4\},\ k\in\{1,2,\ldots,8\}$:
\begin{equation}\label{eq:qjhk0}
  \langle \mathbf{q}_{j},\mathbf{h}_k\rangle=0,\quad \forall k\neq j\ \mathrm{or}\ \forall k\neq j+4.
\end{equation}
Moreover, using (\ref{eq:qjhk0}), we have for $j\in\{5,6,7,8\}$:
\begin{equation}
  \mathbf{r}_j=\mathbf{h}_j-\langle \mathbf{q}_{j-4},\mathbf{h}_j\rangle\mathbf{q}_{j-4}.
\end{equation}
Therefore, using (\ref{eq:hjhk}) and (\ref{eq:qjhk0}), it yields that, for $j\in\{5,6,7,8\}$ and $k\in\{1,2,\ldots,8\}$:
\begin{align}\label{eq:qjhk4}
  \langle \mathbf{q}_{j},\mathbf{h}_k\rangle&=(\langle \mathbf{h}_{j},\mathbf{h}_k\rangle-\langle \mathbf{q}_{j-4},\mathbf{h}_j\rangle\langle \mathbf{q}_{j-4},\mathbf{h}_k\rangle)/\|\mathbf{r}_j\|\nonumber\\
  &=0,\qquad \forall\  k\neq j\ \mathrm{or}\ \forall k\neq j+4.
\end{align}
Combining (\ref{eq:qjhk0}) and (\ref{eq:qjhk4}), it completes the proof.

\emph{Remark}: Theorem 1 indicates that six real-valued information  symbols are uncorrelated, as we designed.
In fact, dividing the four complex symbols $\{s_1,s_2,s_3,s_4\}$ into four groups
$\{s_1^R,s_3^R\}$, $\{s_1^I,s_3^I\}$, $\{s_2^R,s_4^R\}$, and $\{s_2^I,s_4^I\}$,
the two symbols within each group are correlated in the received signal, while symbols of different groups are uncorrelated, as illustrated in Fig.~\ref{fig_R_mat_new}.
This independency permits us to reduce the searching space in the decoding process, which will be presented in the following section.

\subsection{Low complexity ML decoding}
The independency between information symbols shown in Theorem 1 can be exploited to reduce the decoding complexity.
For instance, the received signal $z_1$ does not contain any contribution from information symbols $s_2$ and $s_4$.
It means that the ML solutions of six information symbols $\{s_1,s_3,s_5,s_6,s_7,s_8\}$ can be jointly determined regardless the choices of $\{s_2,s_4\}$.
Similarly, $z_2$ does not have contribution from $s_1$ and $s_3$.
The six information symbols $\{s_2,s_4,s_5,s_6,s_7,s_8\}$ can be decided together without considering the solutions of $\{s_1,s_3\}$.
Hence, the overall detection of eight information can be carried out by a joint detection of four symbols $\{s_5,s_6,s_7,s_8\}$ followed by two independent detections of $\{s_1,s_3\}$ and $\{s_2,s_4\}$ in parallel.
In general, the ML decoding is realized by joint searches of six information symbols which results a complexity of $O(M^6)$.
Note that the parallel detections do not rely on the characteristic of the constellation.
In other words, this complexity reduction is applicable for \emph{arbitrary} constellation scheme.

Furthermore, when the square $M$-QAM is adopted, the detections of real parts and imaginary parts of information symbols can be decoupled.
Rewrite the $16\times16$ real-valued $\mathbf{R}$ matrix by:
\begin{equation}\label{eq:}
    \mathbf{R}=\left [ \begin{array}{*{20}c}
    \mathbf{R}_1 & \mathbf{R}_2 \\
    \mathbf{O} & \mathbf{R}_4 \\
    \end{array} \right],
\end{equation}
where $\mathbf{R}_1$ $\mathbf{R}_2$ and $\mathbf{R}_4$ are $8\times8$ submatrices. Separating the symbol vectors $\mathbf{s}$ and $\mathbf{z}$ in two groups, i.e. $\mathbf{s}^{(1)}=[s_1,s_2,s_3,s_4]^T$, $\mathbf{s}^{(2)}=[s_5,s_6,s_7,s_8]^T$, $\mathbf{z}^{(1)}=[z_1,z_2,z_3,z_4]^T$ and $\mathbf{z}^{(2)}=[z_5,z_6,z_7,z_8]^T$, the ML decoding in (\ref{eq:SD_detection_metric}) is converted into  a conditional detection:
\begin{align}
&\arg\min_{\mathbf{s}\in\boldsymbol\Theta^8}(\|\widetilde{\mathbf{z}}^{(1)}-\mathbf{R}_1\widetilde{\mathbf{s}}^{(1)}-\mathbf{R}_2\widetilde{\mathbf{s}}^{(2)}\|^2+\|\widetilde{\mathbf{z}}^{(2)}-\mathbf{R}_4\widetilde{\mathbf{s}}^{(2)}\|^2)\nonumber \\
 &= \arg\!\!\!\!\min_{\mathbf{s}^{(2)}\in\boldsymbol\Theta^4}\!\!(\|\widetilde{\mathbf{z}}^{(2)}\!-\!\mathbf{R}_4\widetilde{\mathbf{s}}^{(2)}\|^2\!+\!\arg\!\!\!\!\min_{\mathbf{s}^{(1)}\in\boldsymbol\Theta^4}\!\!\|\widetilde{\mathbf{v}}^{(1)}\!-\!\mathbf{R}_1\widetilde{\mathbf{s}}^{(1)}\|^2),\label{eq:cond_dec}
\end{align}
where $\widetilde{\mathbf{v}}^{(1)}=\widetilde{\mathbf{z}}^{(1)}-\mathbf{R}_2\widetilde{\mathbf{s}}^{(2)}$.
In addition, if we take into account the property of $\mathbf{R}_1$ given in Theorem 1, the inner search of four complex symbols $\mathbf{s}^{(1)}$ is simplified to be four independent searches for $\{s_1^R,s_3^R\}$, $\{s_1^I,s_3^I\}$, $\{s_2^R,s_4^R\}$, and $\{s_2^I,s_4^I\}$, respectively.
More precisely, we have:
\begin{align}
&\arg\!\!\!\!\min_{\mathbf{s}^{(1)}\in\boldsymbol\Theta^4}\!\!\|\widetilde{\mathbf{v}}^{(1)}\!-\!\mathbf{R}_1\widetilde{\mathbf{s}}^{(1)}\|^2 =\nonumber \\
&\arg\min_{s_3^R\in\boldsymbol\Psi}((v_1^{R}-R_{1,1} \bar{s}_1^{R}-R_{1,5} s_3^{R})^2+(v_3^R-R_{5,5} s_3^R)^2) \nonumber \\
+&\arg\min_{s_3^I\in\boldsymbol\Psi}((v_1^{I}-R_{2,2}\ \bar{s}_1^{I}-R_{2,6}\ s_3^{I})^2+(v_3^I-R_{6,6}\ s_3^I)^2) \nonumber \\
+&\arg\min_{s_4^R\in\boldsymbol\Psi}((v_2^{R}-R_{3,3} \bar{s}_2^{R}-R_{3,7} s_4^{R})^2+(v_4^R-R_{7,7} s_4^R)^2) \nonumber \\
+&\arg\min_{s_4^I\in\boldsymbol\Psi}((v_2^{I}-R_{4,4}\ \bar{s}_2^{I}-R_{4,8}\ s_4^{I})^2+(v_4^I-R_{8,8}\ s_4^I)^2), \label{eq:parallel_dec}
\end{align}
where $\boldsymbol\Psi$ is the set of $\sqrt{M}$-PAM constellation symbols; $R_{j,k}$ represents the $(j,k)$th element of the $\mathbf{R}$ matrix; $\bar{s}_1^{R}$, $\bar{s}_1^{I}$, $\bar{s}_2^{R}$ and $\bar{s}_2^{I}$ are the $\sqrt{M}$-PAM constellation symbols that minimize the ML decoding metrics given $s_3^{R}$, $s_3^{I}$, $s_4^{R}$ and $s_4^{I}$, respectively, and can be obtained via simple hard decisions:
\begin{align}
s_1^R&=Q\Big( \frac{v_1^R-R_{1,5}\ s_3^R}{R_{1,1}}\Big),\quad s_1^I=Q\Big( \frac{v_1^I-R_{2,6}\ s_3^I}{R_{2,2}}\Big), \\
s_2^R&=Q\Big( \frac{v_2^R-R_{3,7}\ s_4^R}{R_{3,3}}\Big), \quad s_2^I=Q\Big( \frac{v_2^I-R_{4,8}\ s_4^I}{R_{4,4}}\Big),
\end{align}
where $Q(x)$ is the hard decision function that returns the $\sqrt{M}$-PAM symbol which is closest to the given value $x$.
It can be seen that the parallel detections in (\ref{eq:parallel_dec}) are realized by searches over $\sqrt{M}$-PAM symbols, resulting a complexity of $O(\sqrt{M})$.

Combining (\ref{eq:cond_dec}) and (\ref{eq:parallel_dec}), the ML decoding requires an overall detection complexity of $O(M^{4.5})$ when the square QAM constellations are used.
The decoding complexities of some state-of-the-art fast-decodable STBCs are given in Table~\ref{tbl:det_complexity}.
It can be seen that the proposed code is among the least complex ones.

\subsection{Optimization of minimum determinants}

\begin{table}[t]
\centering
\begin{threeparttable}
\caption{Comparison of minimum determinants and ML decoding complexities of fast decodable STBCs}
\label{tbl:det_complexity}
\footnotesize
\renewcommand\arraystretch{1.2}
\begin{tabular}{|c|c|c|c|}
\hline
\multirow{2}{*}{\textbf{STBC}} & \multirow{2}{*}{\textbf{Min determinant}} & \multicolumn{2}{c|}{\textbf{ML decoding complexity}}\\ \cline{3-4}
& &any QAM & square QAM \\ \hline
Proposed code & $10.24$ & $O(M^6)$ & $O(M^{4.5})$ \\ \hline
DjABBA~\cite{hottinen2004precoder}& $0.8304$ \tnote{$\dagger$} & $O(M^7)$ & $O(M^{6})$\\ \hline
Srinath-Rajan~\cite{srinath2009low}& $10.24$ & $O(M^5)$ & $O(M^{4.5})$\\ \hline
3D MIMO~\cite{nasser20083d} & $0.0318$ & $O(M^6)$ & $O(M^{4.5})$\\ \hline
IFS rate 2~\cite{ismail2013new} & $0.0076$ & $O(M^5)$ & $O(M^{4.5})$\\ \hline
BHV~\cite{biglieri2009fast} & $0$ & $O(M^6)$ & $O(M^{4.5})$\\ \hline
\end{tabular}
{\footnotesize
\begin{tablenotes}
\item [$\dagger$] Using the best rotation $\rho=\cos^{-1}(0.8881)$~\cite{hottinen2004precoder} known in the literature.
\end{tablenotes}
}
\end{threeparttable}
\end{table}

It is well known that in the STBC design the minimum determinant of the codeword difference matrix should be maximized to achieve higher coding gain which consequently leads to better pairwise error probability (PEP) performance~\cite{tarokh1998space}.
In the literature, constellation rotation is a common way to maximize the coding gain~\cite{biglieri2009fast,srinath2009low,ismail2013new}.
The rotated constellation brings additional diversities between in-phase and quadrature components of the signal.
As far as the proposed codeword is concerned, the rotation angle $\rho$ actually performs the constellation rotation
and should be optimized in order to maximize the minimum determinant of the codeword difference matrix.
It is worth noting that the choice of $\rho$ does not affect the fast decodability of the codeword.

We propose to use the rotation angle $\rho = \tan^{-1}(\frac{1+\sqrt{5}}{2})$, a value originated from the Golden number, in the new codeword (\ref{eq:newDjABBA}).
It has been shown in theory that a constellation rotation with Golden number actually gives best performance~\cite{belfiore2005golden,srinath2009low}.
In this work, it is proved through exhaustive search that the minimum determinant of the proposed codeword with the ``Golden rotation'' is $10.24$ for unnormalized QAM constellations. This is the highest value for the $4\times2$ full-rate STBCs reported in the literature~\cite{srinath2009low,ismail2013new}.
A comparison of the minimum determinant of different $4\times2$ full-rate STBCs is given in Table~\ref{tbl:det_complexity}.
It can be seen that the proposed code has the same determinant as the Srinath-Rajan code and has higher determinant than other STBCs.
Finally, it is worth noting that the non-zero minimum determinant also indicates that the proposed new code achieves full-diversity for the four-transmit-antenna MIMO transmissions.

\section{Simulation Results}
\label{sec:simu}
\begin{figure}[!t]
\centering
\includegraphics[width=3.3in]{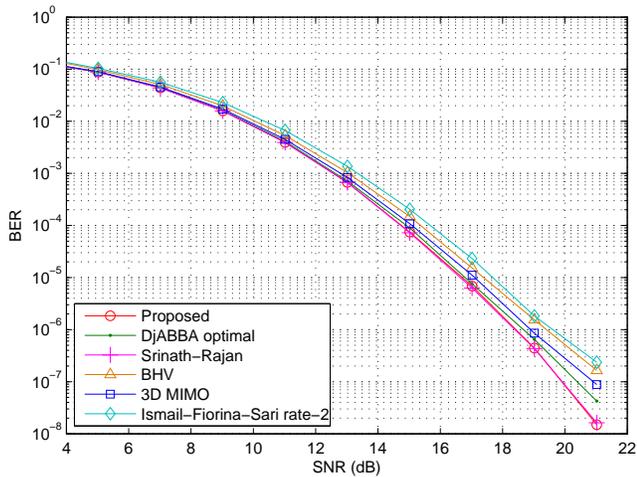}
\caption{BER performance comparison of different $4\times2$ STBCs, QPSK, i.i.d. Rayleigh channel.}
\label{fig_BER_QPSK}
\end{figure}

\begin{figure}[!t]
\centering
\includegraphics[width=3.3in]{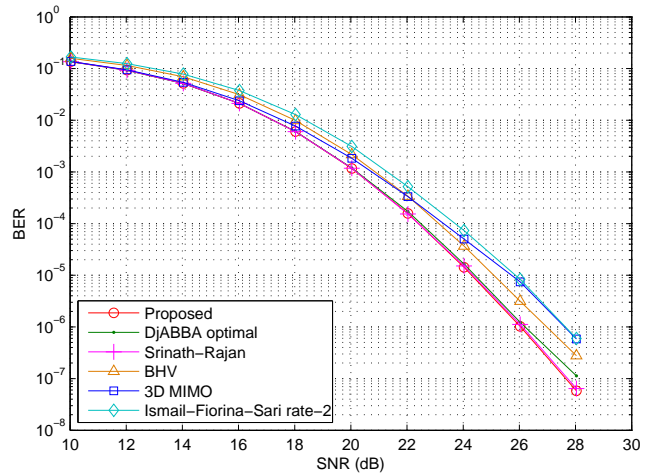}
\caption{BER performance comparison of different $4\times2$ STBCs, 16QAM, i.i.d. Rayleigh channel.}
\label{fig_BER_16QAM}
\end{figure}

Fig.~\ref{fig_BER_QPSK} and Fig.~\ref{fig_BER_16QAM} present the bit error rate (BER) performance of different STBCs  with QPSK and 16QAM, respectively.
The channel model used in the simulation is the $4\times2$ MIMO channel with i.i.d. Rayleigh flat fading coefficients for all channel links.
The ML decoder is used to decode the received MIMO signal.
No channel coding scheme is used in the simulation to give the comparison of ``pure'' performance of the STBCs
The rotation angle for the DjABBA code is chosen as $\rho=\cos^{-1}(0.8881)$ which is the best value known in the literature~\cite{hottinen2004precoder}.

From the figures, it can be seen that the proposed new code provides the best BER performance among all STBCs considered in the comparison with both QPSK and 16QAM.
This can be explained by the fact that the proposed code has the highest minimum determinant.
In particular, the proposed new code achieves the same performance as the Srinath-Rajan code which has the same coding gain, while it outperforms other state-of-the-art STBCs such as 3D MIMO code, BHV code and IFS rate-2 code.
Moreover, the new code performs better than the DjABBA code. It means that the proposed Golden rotation angle provides a better performance than the best proposal  existing in the literature.
In general, the simulation results prove the advantage of the proposed code in terms of the superior BER performance.

\section{Conclusion}
\label{sec:conclusion}

In this work, we propose a new fast-decodable full-rate full-diversity STBC for $4\times2$ MIMO systems. The new code requires a ML decoding complexity of $O(M^{4.5})$ which is the least among all full-rate $4\times2$ STBCs in the literature. Moreover, with the proposed Golden rotation angle, the new code the possesses highest coding gain which provides a better PEP performance compared with other state-of-the-art STBCs. This is proved by the simulation results which show that the proposed code has superior BER performance.

\section*{Acknowledgment}

The authors would like to thank the support of French ANR project ``Mobile Multi-Media (M3)'' and ``P\^ole Images \& R\'eseaux''.

\ifCLASSOPTIONcaptionsoff
  \newpage
\fi



%

\end{document}